\documentclass[aps,pre,reprint,onecolumn,12pt,showkeys]{revtex4-2}
\usepackage[a4paper,left=25mm,top=25mm,textwidth=160mm,textheight=249.7mm]{geometry}

\usepackage{color}
\usepackage{mathtools}

\usepackage{hyperref}
\hypersetup{ colorlinks,
linkcolor=blue,
filecolor=blue,
urlcolor=blue,
citecolor=blue}

\usepackage{amsmath}
\usepackage{amssymb}
\usepackage{wasysym}
\usepackage{mathtools}
\usepackage{xcolor}

\newcommand{\B}{\color{blue}}

\begin{document}

%\title{Phase diagram of YK model with exchange diffusion on square %and hexagonal lattices}

\title{Phase Diagrams of the YK Surface-Reaction Model on 2D lattices with Exchange Diffusion}

\author{Henrique A. Fernandes\textsuperscript{1}}
\altaffiliation{\textsuperscript{}{\B Corresponding author: hafernandes@ufj.edu.br}}
\author{Roberto da Silva\textsuperscript{2}}
\author{Paulo F. Gomes\textsuperscript{1,3}}
%\author{Henrique A. Fernandes\textsuperscript{1}, Roberto da Silva\textsuperscript{2},   Paulo F. Gomes\textsuperscript{1,3}}
%\altaffiliation{\textsuperscript{}{\B Corresponding author: hafernandes@ufj.edu.br}}

\affiliation{1 - Instituto de Ciências Exatas e Tecnológicas, Universidade Federal de Jata{\'i}, BR 364, km 192, 3800 - CEP 75801-615, Jata{\'i}, Goi{\'a}s, Brazil}
\affiliation{2 - Instituto de F{\'i}sica, Universidade Federal do Rio Grande do Sul, Av. Bento Gon\c{c}alves, 9500 - CEP 91501-970, Porto Alegre, Rio Grande do Sul, Brazil}
\affiliation{3 - Faculdade de Ci{\^e}ncias e Tecnologia, Universidade Federal de Goi{\'a}s, Estrada Municipal, Bairro Fazenda Santo Ant{\^o}nio, CEP 74971-451, Aparecida de Goi{\^a}nia, Goi{\'a}s - Brasil}

\begin{abstract}

In this work, we investigate the phase diagrams of the Yaldram and Khan catalytic surface model on square and hexagonal lattices when exchange diffusion is allowed for carbon monoxide (CO) and nitrogen (N) atoms. To reach our goal, we carried out steady-state Monte Carlo (MC) simulations over $4\times 10^5$ points, for both lattices, in order to obtain a framework of the steady reactive state of the model for different values of the nitric oxide dissociation rate, $r_{NO}$. The results show the emergence of steady reactive state for certain values of $r_{NO}$ and of exchange diffusion rate $x$ when the simulations take place on square lattices. Our findings on the hexagonal lattice also show that the diffusion of these species favors the appearance of the active phase for values of $r_{NO}$ lower than that found for the standard model. In addition, we observed that the system possesses a continuous phase transition and a discontinuous one separating the active phase from absorbing states for both lattices, except for $r_{NO}=1$ in which the continuous phase transition is destroyed and a steady reactive state emerges from the beginning since very small values of $x$.  


\end{abstract}

\maketitle

\section{Introduction}
\label{introduction}

Catalytic reactions on surfaces have played a central role in both theoretical and experimental studies, in part due to their industrial and technological importance, with an enormous number of applications, and also because they consist of non-equilibrium systems which can exhibit complex behaviors, as for example, oscillations, chaotical behavior, self-organization and non-equilibrium kinetic phase transitions \cite{Bond1987, Christmann1991, Imbihl1995, Ziff1986, Ehsasi1989, Yaldram1991, Block1993, Marro1999, Baxter2002, Loscar2003}.

Monte Carlo (MC) simulations have become one of the main tools to help understand these complex out-of-equilibrium processes. As shown by several authors, in addition to overcome the difficulties found in analytical studies, results obtained from computational simulations of lattice-gas models for heterogeneous catalytic reactions capture, in a qualitative way, part of the complexity observed in experiments.

One of the most studied catalytic surface models was devised in 1986 by R. M. Ziff, E. Gulari and Y. Barshad \cite{Ziff1986} with the intention of mimicking $\text{CO}-\text{O}_2$ catalytic reactions. In that work, the surface was represented by square lattices and the reactions were very straightforward. Despite its simplicity, the model presented a rich phase diagram, with continuous and discontinuous phase transitions separating two poisoned states from the active one where there exists sustaintable production of carbon dioxide, $\text{CO}_2$, molecules. Since then, the well-known ZGB model has been extensively studied by several author through different approaches and techniques as can be seen, for instance, in Refs. \cite{Meakin1987, Dickman1986, Fischer1989, Tome1993, Dumont1990, Albano1992, Kaukonen1989, Jensen1990, Brosilow1992a, Buendia2013, Buendia2015, chan2015, Grandi2002, Hoenicke2000, Buendia2012, Hoenicke2014, Satulovsky1992, Albano1990, Brosilow1993, Fernandes2016primeiroZGB, RdasilvadifusionZGB2018, Fernandesdesopsion, Vilela2020, Fernandes2025}.

A few years after the advent of the ZGB model, another prominent catalytic surface model was proposed by K. Yaldram and M. A. Khan in order to take into consideration the oxidation of CO and the reduction of NO \cite{Yaldram1991} molecules. In that work, the authors considered two different regular lattices as catalytic surfaces: square and hexagonal lattices. They showed that there is no active phase when the kinetics ocurrs on square lattices, for any value of the dissociation rate $r_{\text{NO}}$, the rate in which the NO molecule dissociates into N and O atoms on the surface. Therefore, the states are poisoned and there is no production of $\text{CO}_2$ and $\text{N}_2$ molecules. The second considered surface was the hexagonal (triangular) lattice and, in that case, the system presented a very interesting phase diagram, as for the ZGB model, with continuous and discontinuous phase transitions for $r_{\text{NO}}>0.8$, as well as a steady reactive window which enlarges from this point until $r_{\text{NO}}=1.0$.

Following the seminal work, the model proposed by Yaldram and Khan (the YK model) 
has been extensively studied through other techniques and considering different processes, such as the inclusion of impurities on the surface \cite{Ahmad2007, Hernandez2022}, the diffusion \cite{Aida1999, Luque2004, Avalos2006, Loscar2003, Khan2002, Khan1994, Dickman1999, Kortluke1996} and desorption \cite{Khan2002, Diaz2018, Meng1994, Hernandez2022, Diaz2017, Kortluke1996} of some particles, as well as by taking into account complex networks as catalytical surfaces \cite{Albano1990, Gomes2025}.

In the present work, we revisit the YK model in order to study the influence of the exchange diffusion of CO molecules and N atoms on the phase diagram of the model in both regular square and hexagonal lattices. To achieve our goal, we performed steady-state Monte Carlo (SSMC) simulations for several values of the dissociation rate, $r_\text{NO}$, and exchange diffusion rate.

In next section, we describe the model, discuss the implementation of the exchange diffusion of particles, as well as the procedure adopted in our numerical simulations. Our main results for both catalytic surfaces are presented in Sec. \ref{results} and, finally, our conclusions are summarized in Section \ref{conclusions}.

\section{The model and the simulation method} \label{model}

The YK model was designed in 1991 \cite{Yaldram1991} in order to mimic the NO-CO catalytic reaction on two distinct surfaces. They had shown, by means of SSMC simulations, that there is no active phase for square surfaces no matter the value of the dissociation rate $r_{NO}$, the rate at which the nitric oxide (NO) molecule dissociates into nitrogen (N) and (O) atom soon after being adsorbed on the surface. On the other hand, their results support the presence of active phase for hexagonal surfaces, with sustaintable production of nitrogen ($\text{N}_2$) and carbon dioxide ($\text{CO}_2$) molecules for certain values of $r_{NO}$ and $y$, the adsorption rate of carbon monoxide (CO) molecules, i.e., the rate at which a CO molecule impinges the catalytic surface.

The chemical reaction can be expressed by
\begin{equation*}
\text{NO}+\text{CO} \longrightarrow \frac{1}{2}\text{N}_2+\text{CO}_2.
\end{equation*}

The processes involved in the dynamics of the model follows the Langmuir-Hinshelwood mechanism and range from the adsorption of molecules on the surface, the dissociation of NO molecules adsorbed on it, to the production and desorption of $\text{N}_2$ and $\text{CO}_2$ molecules from the surface. They can be summarized as follows:
\begin{align}
\text{CO}(g)+V &\overset{y}{\longrightarrow} \text{CO}(a), \label{eq:co_ad}  \\
\text{NO}(g)+V &\overset{1-y}{\longrightarrow} \text{NO}(a), \label{eq:no_ad}  \\
\text{NO}(g)+2V &\overset{r_{\text{NO}}}{\longrightarrow} \text{N}(a) + \text{O}(a), \label{eq:n_o_ad} \\
\text{NO}(a)+\text{N}(a) &\overset{1}{\longrightarrow} \text{N}_2(g)+\text{O}(a)+V, \label{eq:n2_g_o_ad} \\
\text{N}(a)+\text{N}(a) &\overset{1}{\longrightarrow} \text{N}_2(g)+2V, \label{eq:n2_g} \\
\text{CO}(a)+\text{O}(a) &\overset{1}{\longrightarrow} \text{CO}_{2}(g) + 2V, \label{eq:co2_des}
\end{align}
where $g$ stands for molecules (CO and NO) in the gas phase, $a$ is related do atoms or molecules adsorbed on the surface, and $V$ represents vacant sites on the surface. The model possesses two parameters: the adsorption rate of CO molecules, $y$, and the dissociation rate of NO molecules, $r_{NO}$. As shown in Eq. (\ref{eq:co_ad}), a CO molecule in the gas phase is chosen to impinge the surface with a rate $y$, being absorbed if it hits a vacant site. At a rate $1-y$, it is a NO molecule in the gas phase that is chosen to collide with the surface, and this process is represented by Eq. (\ref{eq:no_ad}). In this case, the NO molecule may or may not dissociate into N and O atoms, depending on the value of the dissociation rate, $r_{\text{NO}}$. Equation (\ref{eq:n_o_ad}) shows that the dissociation occurs with that rate and the atoms are adsorbed on the surface whenever two neighboring sites, chosen at random, are empty. On the other hand, if the dissociation does not occur, the NO molecule is adsorbed in a single site of the surface if the chosen site is vacant. 

As we can observe, if either chosen site is occupied by an atom/molecule during any adsorption process, the trial ends and the CO or NO molecule returns to the gas phase. Otherwise, whenever an atom/molecule is adsorbed on the surface, its neighboring sites must be checked in order to take into account possible reactions with production of $\text{N}_2$ or $\text{CO}_2$ molecules. Equation (\ref{eq:n2_g_o_ad}) shows that when a NO molecule is adsorbed and there is, at least, one neighboring N atom, a $\text{N}_2$ molecule is produced and desorbs from the surface, leaving behind the O atom at the same site where the NO molecule was before the catalytic reaction, as well as a vacant site where the N atom was previously. Once this reaction occurs, the neighboring sites of the N and O atoms left on the surface must also be analysed in order to check for other possible reactions, as shown in Eqs. (\ref{eq:n2_g}) and (\ref{eq:co2_des}). In the first case, the N atom reacts with another neighboring N atom producing a $\text{N}_2$ molecule and, in the second one, the O atom reacts with a neighboring CO molecule producing a $\text{CO}_2$. In both scenarios, the molecules desorb from the surface, going to the gas phase and, therefore, leaving two empty sites on the surface. Equation (\ref{eq:co2_des}) also stands for the adsorption process of the CO molecule, i.e., when such a molecule is adsorbed on the surface, a random check of its neighboring sites is performed in order to look for O atoms. If the search is successful, a $\text{CO}_2$ molecule is immediately formed and desorbed from the surface, leaving two vacant sites on it.

In the present work, we study the influence of exchange diffusion of N atoms and CO molecules on the phase diagram of the standard YK model presented above. By exchange diffusion, we mean that the particles swap places rather than moving only when they find a neighboring vacant site.

Although adding this process to the model is very simple, it produces big changes in its phase diagram, mainly when the simulation occurs on square lattices. The process occurs as follows: If a site $i$, chosen at random, is vacant, the attempt of adsorption of CO or NO molecules is performed, as usual. However, if the chosen site is occupied by a N atom or a CO molecule, the diffusion may occur according to the diffusion rate $x$, as shown in the following equations:
\begin{align}
\text{CO}_i+\text{N}_j &\overset{x}{\longrightarrow} \text{N}_i+\text{CO}_j, \label{eq:co_n_dif}  \\
\text{N}_i+\text{CO}_j &\overset{x}{\longrightarrow} \text{CO}_i+\text{N}_j, \label{eq:n_co_dif}  \\
\text{CO}_i+V_j &\overset{x}{\longrightarrow} V_i+\text{CO}_j, \label{eq:co_v_dif} \\
\text{N}_i+V_j &\overset{x}{\longrightarrow} V_i+\text{N}_j, \label{eq:n_v_dif}
\end{align}
where $j$ stands for the neighboring site of $i$, also chosen at random.

In that case, if a neighboring site is also occupied by a N atom or a CO molecule, or is vacant, the exchange diffusion occurs, i.e., the species involved in the process change place, otherwise, the trial ends. Of course, if both chosen sites ($i$ and $j$) are occupied by the same species, the trial ends since there is no reason for the diffusion process to occur. On the other hand, if the diffusion is carried out, their neighboring sites are checked following Eqs. (\ref{eq:n2_g_o_ad})-(\ref{eq:co2_des}).

The quantities of interest are the densities of species adsorbed on the surface, as well as the density of vacant ($V$) sites, and the densities of $\text{CO}_2$ and $\text{N}_2$ molecules produced during the processes given by Eqs. (\ref{eq:n2_g_o_ad})-(\ref{eq:co2_des}). They are given by
\begin{equation}
\rho_s=\frac{N_s}{N},
    \label{eq:densities}
\end{equation}
where $s$ stands for the species considered in the calculation, $N_s$ is 
the number of the species $s$ on the surface or produced during one MC step, and $N$ is the total number of sites on the lattice.

In our SSMC simulations, we considered regular square and hexagonal lattices of linear size $L=64$ and the averages were calculated after discarding the first $3\times 10^5$ steps needed for the system to reach the steady state regime. 

We also took into consideration the following values of the dissociation rate, $r_\text{NO}$: 0.700, 0.750, 0.800, 0.850, 0.900, 0.950, and 1.00. In order to obtain a framework of the phase diagrams of the model, we varied the adsorption rate, $y$, from 0.001 to 0.400, with $\Delta y=10^{-3}$, and the exchange diffusion rate, $x$, from 0 to 1.000, also with $\Delta x=10^{-3}$ producing a total of $4\times 10^5$ points which were used in the construction of the phase diagrams. As we show in the next section, these parameters allows us to observe the very beginning of the active state, as well as its behavior, of both catalytic surfaces.

\section{Results and discussion}
\label{results}

As shown in Sec. \ref{introduction}, diffusion processes in YK model have already been considered in previous works, showing the emergence of active states when certain particles are free to move on a square lattice for specific values of $r_{\text{NO}}$. However, in this work, we consider a different diffusion process, where neighboring CO molecules and N atoms adsorbed on the surface can move by exchanging places with each other or, as usual, they can move to a neighboring vacant site, both with a diffusion rate $x$.

In our work, we are particularly interested in phase diagrams which present steady reactive states (active phases) with the presence of vacant sites and the consequent production of $\text{CO}_2$ and $\text{N}_2$ molecules. Then, by looking into the density $\rho_V$ as function of $y$ and $x$ we will be able to observe the existence (or not) of active phases in the model.

For both square and hexagonal lattices, there is no active phase when the dissociation rate $r_{NO}$ is equal to 0.700, regardless of the value of the exchange diffusion rate, $x$. However, for $r_{NO} \geq 0.750$, there exist steady reactive states for both catalytic surfaces, meaning that the emergence of active phases occurs between $0.700 < r_{NO} < 0.750$.

\begin{figure} [!htbp] %h or !htbp
\begin{center}
\includegraphics[width = 5.0 in]{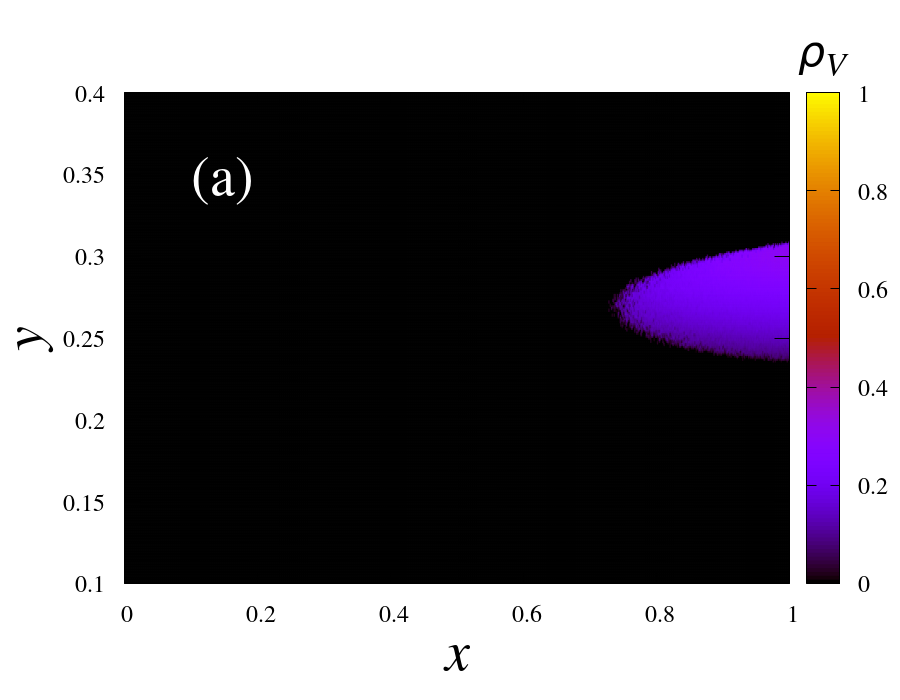}
\includegraphics[width = 5.0 in]{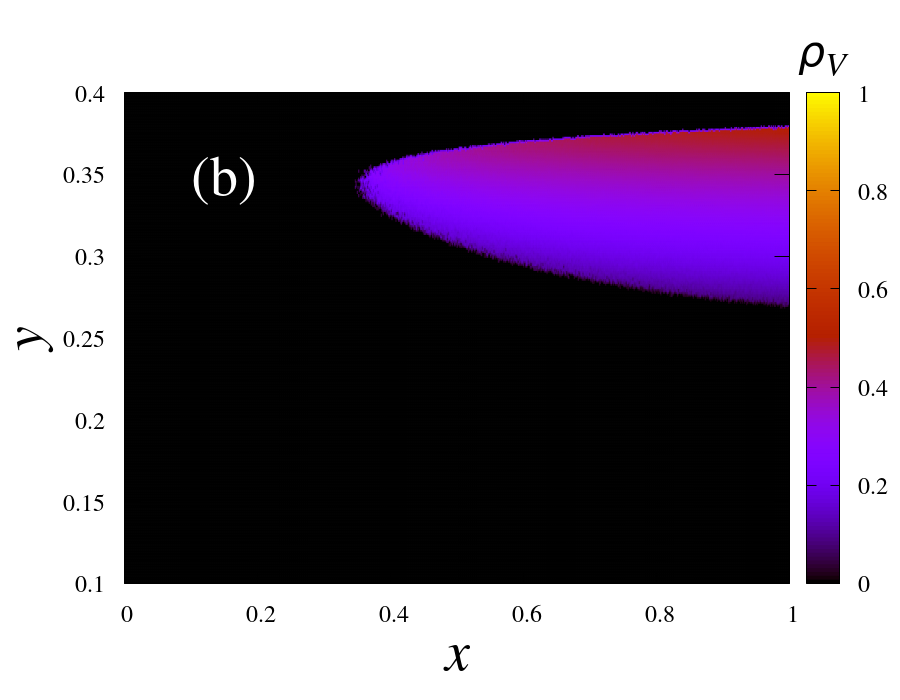}
\caption{Phase diagram of the YK model for (a) square and (b) hexagonal lattices for $r_{NO}=0.750$. Each figure shows $4\times 10^5$ points in $(x,\ y)$ space, each one representing the density of vacant sites, $\rho_V$. The black color stands for poisoned states, i.e., $\rho_V=0$.}
\label{fig:res_vaz_rno_0.750}
\end{center}
\end{figure}

Figure \ref{fig:res_vaz_rno_0.750} shows the density of vacant sites $\rho_V$ as a function of the adsorption rate $y$ and the diffusion rate $x$ for the square (a) and hexagonal (b) lattices. In these figures, black points ($\rho_V=0$) mean that the system is in a poisoned (absorbing) state and points with different colors ($\rho_V\neq 0$) mean that the system is in an active phase. 

In Fig. \ref{fig:res_vaz_rno_0.750}(a), we can see that an active phase emerges for higher values of $x$ ($ \simeq 0.8$) and the reactive window, that is the range in $y-$space, for a given $x$, in which the active phase occurs, increases monotonically with $x$. Figure \ref{fig:res_vaz_rno_0.750}(b) shows that the behavior of the phase diagram for the hexagonal lattice is similar to that of the square one, but the active phase starts earlier, for $x \simeq 0.4$. This figure also shows that, for $x=0$, i.e., when the standard YK model is recovered, there is no active phase as predicted by several authors. In fact, as it is well known, the active phase of the standard YK model begins for $r_{NO}$ close to 0.8. In summary, these figures show us that the exchange diffusion of CO molecules and N atoms on the surfaces damatically modifies the model at square lattices and enables the existence of steady reactive states for smaller values of $r_{NO}$ for hexagonal lattices.

\begin{figure} [!htbp] %h or !htbp
\begin{center}
\includegraphics[width = 5.0 in]{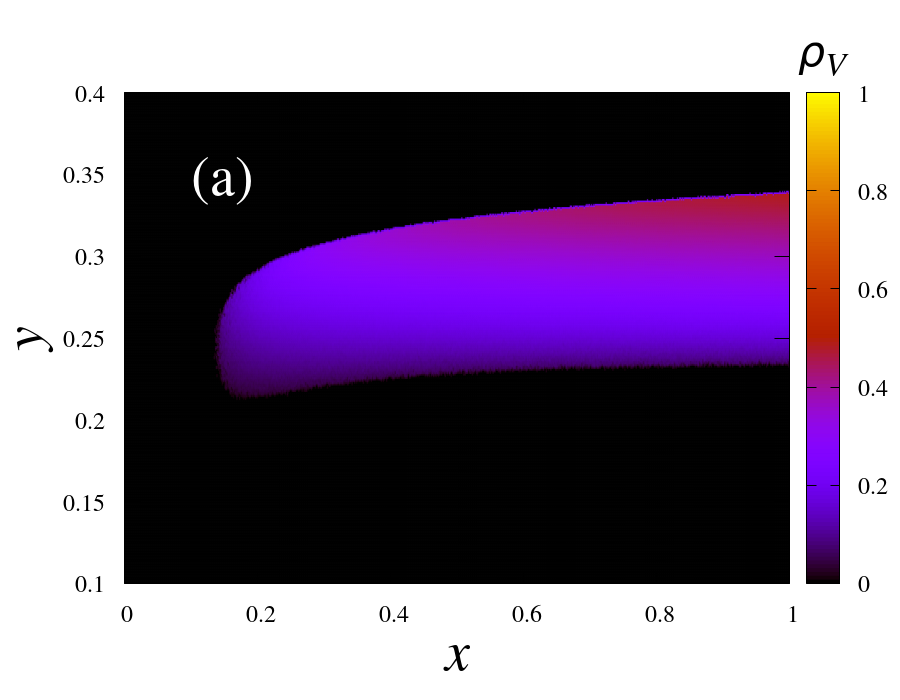}
\includegraphics[width = 5.0 in]{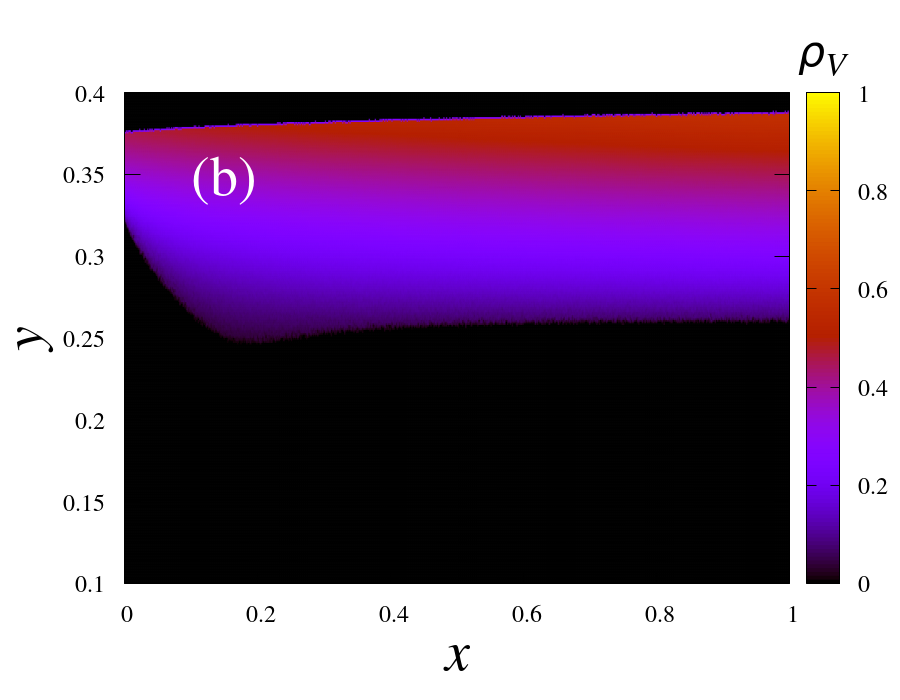}
\caption{Phase diagram of the YK model for (a) square and (b) hexagonal lattices for $r_{NO}=0.800$. Each figure shows $4\times 10^5$ points in $(x,\ y)$ space, each one representing the density of vacant sites, $\rho_V$. Black color stands for poisoned states, i.e., $\rho_V=0$.}
\label{fig:res_vaz_rno_0.800}
\end{center}
\end{figure}

This assertion can be observed in Fig. \ref{fig:res_vaz_rno_0.800}, which presents the phase diagram of the model for $r_{NO}=0.800$. First, Fig. \ref{fig:res_vaz_rno_0.800}(a) shows that the steady reactive phase for the square lattice grows rapidly and now starts for $x\simeq 0.2$.
On the other hand, the active phase for the hexagonal lattice ranges from $0 \leq x \leq 1$, i.e., the system possesses steady reactive state for the entire range of diffusion rates. As we can see, the smallest reactive window occurs for the standard (static) model, $x=0$, where $0.325 \leq y \leq 0.376$, and then it increases rapidly until $x\simeq 0.17$, remaining almost constant from this value to $x=1$.

\begin{figure} [!htbp] %h or !htbp
\begin{center}
\includegraphics[width = 5.0 in]{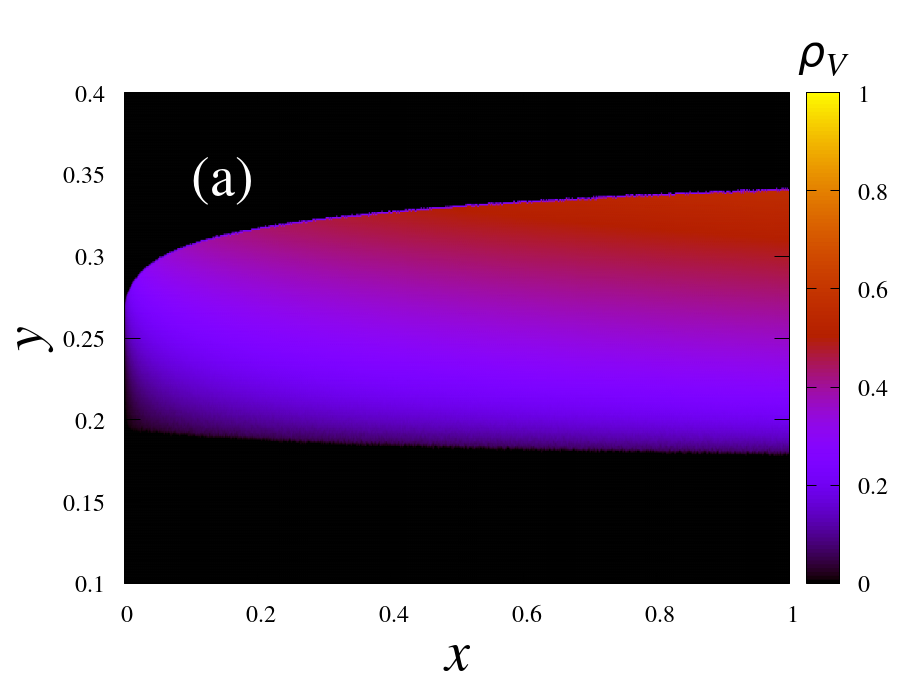}
\includegraphics[width = 4.6 in]{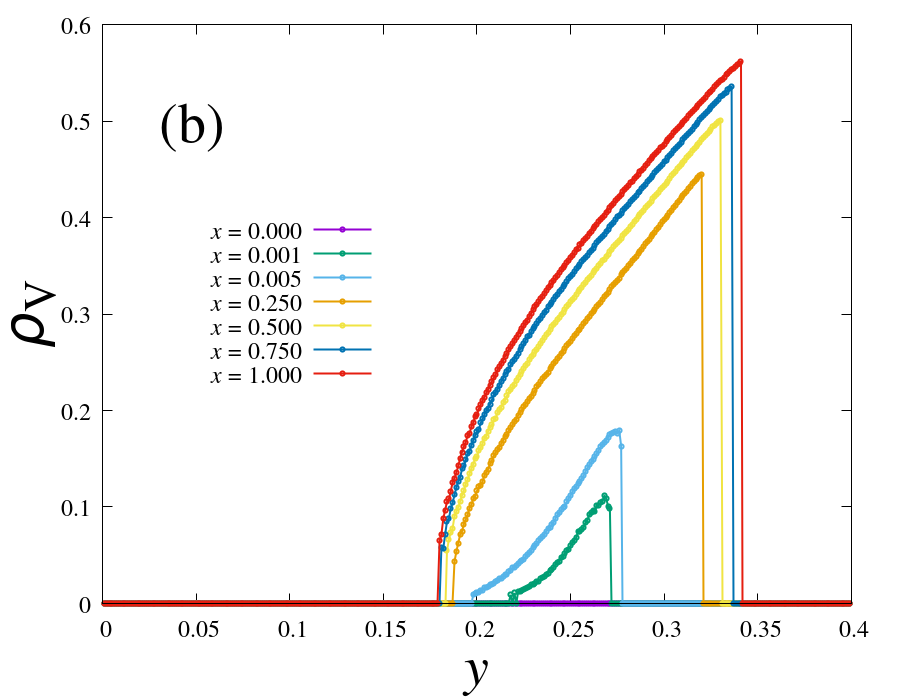}
\caption{(a) Phase diagram of the YK model for square lattices for $r_{NO}=0.900$. (b) Density of vacant sites, $\rho_V$, as function of the adsorption rate $y$ for square lattices and different values of $x$. The densities were estimated by considering $N=2\times 10^4$ MC steps.}
\label{fig:res_vaz_rno_0.900}
\end{center}
\end{figure}

Similar results are obtained for $r_{NO}=0.850$, 0.900, and 0.950, and therefore, are not presented in this paper. Instead, it would be more interesting to look into the lines that outline the steady reactive state, i.e., the line separating the absorbing states from the active phase, in order to observe the order of the phase transitions when the exchange diffusion is present. To proceed with this analysis, we took into consideration the results obtained for $r_{NO}=0.900$, $0.001 \leq y \leq 0.400$, and $x=0$, 0.001, 0.005, 0.250, 0.500, 0.750 and 1.000, as shown in Fig. \ref{fig:res_vaz_rno_0.900} for the square lattice.

Figure \ref{fig:res_vaz_rno_0.900}(a) shows that the active phase emerges even for very small diffusion values. In fact, there is no active phase only at $x=0$, i.e., for the standard YK model, as it should have. The botton line separating the absorbing state from the active one is almost constant while the upper line separating the active phase from the absorbing state grows monotonically since de beginning. In other words, on the contrary of the bottom line, the upper line shows that the phase transition points are sensitive to changes in the diffusion rate, as can also be seen in Fig. \ref{fig:res_vaz_rno_0.900}(b), which presents the results for the density of vacant sites $\rho_V$ as function of $y$ for several values of $x$. As we can note, there is active phase even for $x=0.001$, which possesses a small reactive window, starting at a continuous phase transition and ending at a discontinuous one. This observation remains valid for all values of $x>0$, showing that the inclusion of exchange diffusion in square lattices recover the phase transitions found in hexagonal ones.

\begin{figure} [!htbp] %h or !htbp
\begin{center}
\includegraphics[width = 3.1 in]{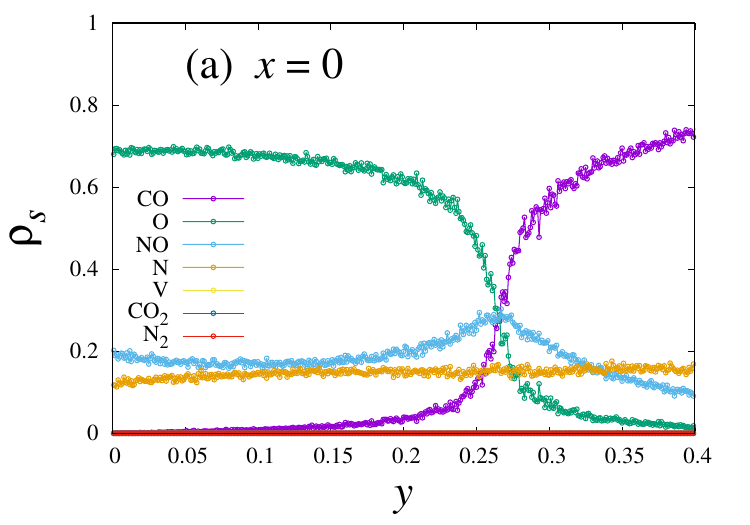}
\includegraphics[width = 3.1 in]{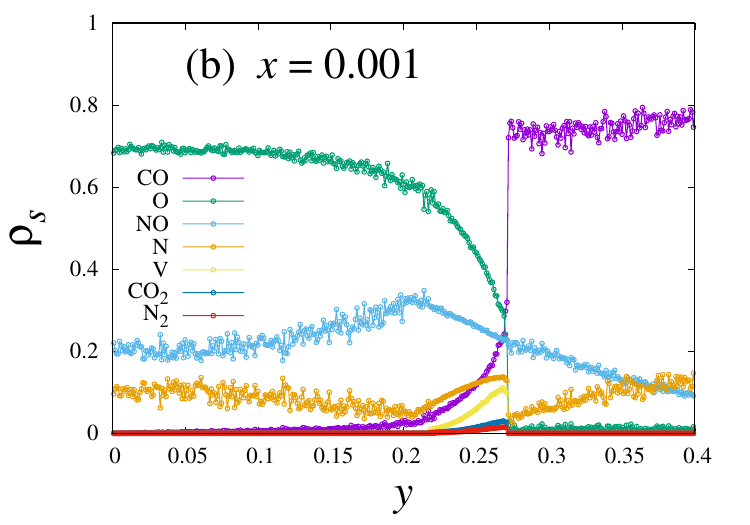}
\includegraphics[width = 3.1 in]{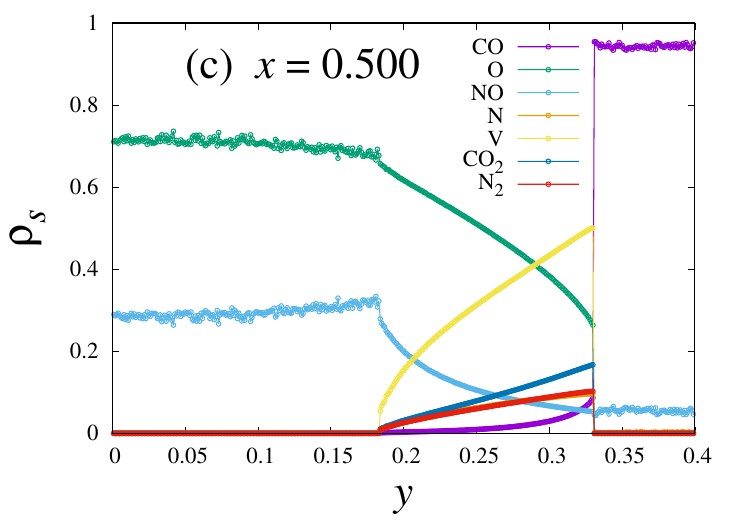}
\includegraphics[width = 3.1 in]{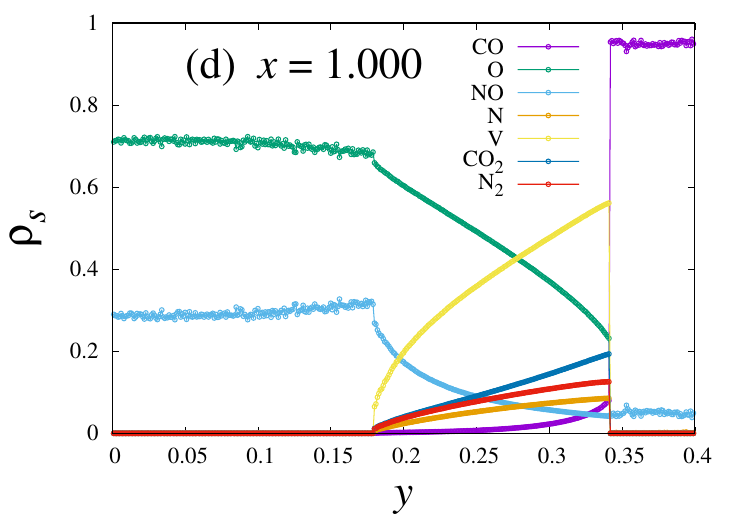}
\caption{Densities $\rho_s$ of the species presented in the YK model for $r_{NO}=0.900$ as function of the adsorption rate $y$ for square lattices and different values of $x$: (a) $x=0$, (b) $x=0.001$, (c) $x=0.500$, and (d) $x=1.000$. The densities were obtained with $N=2\times 10^4$ MC steps.}
\label{fig:snapshot_square}
\end{center}
\end{figure}

In Fig. \ref{fig:snapshot_square}, we show some snapshots of the densities of all species for specific values of $x$: (a) 0, (b) 0.001, (c) 0.500, and (d) 1.000, for $r_{NO}=0.900$, when the model is simulated on square lattices.
Figure \ref{fig:snapshot_square}(a) shows that there is no steady reactive state and, therefore, there is no sustainable production of $\text{CO}_2$ and $\text{N}_2$ molecules. In that case we are dealing with the standard model on square lattices ($x=0$) and this result follows those found by Yaldram and Khan in 1991 \cite{Yaldram1991}. On the other hand, Figs. \ref{fig:snapshot_square}(b), (c), and (d) show that the diffusion of some particles in square lattices ($x\neq 0$) also produces one active phase in between two poisoned states, as shown in previous figures. As one can see, the active phase is surrounded by a second-order phase transition point on one side and a first-order transition point on the other. Similar results were found in Ref. \cite{Yaldram1991} when the catalytic surface is simulated through a regular hexagonal lattice. 

\begin{figure} [!htbp] %h or !htbp
\begin{center}
\includegraphics[width = 5.0 in]{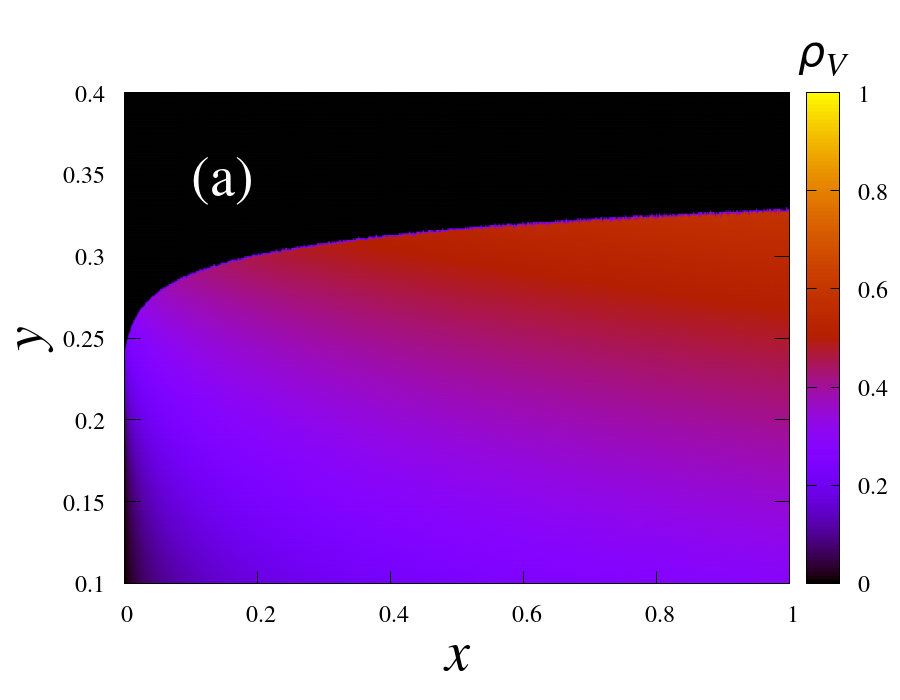}
\includegraphics[width = 5.0 in]{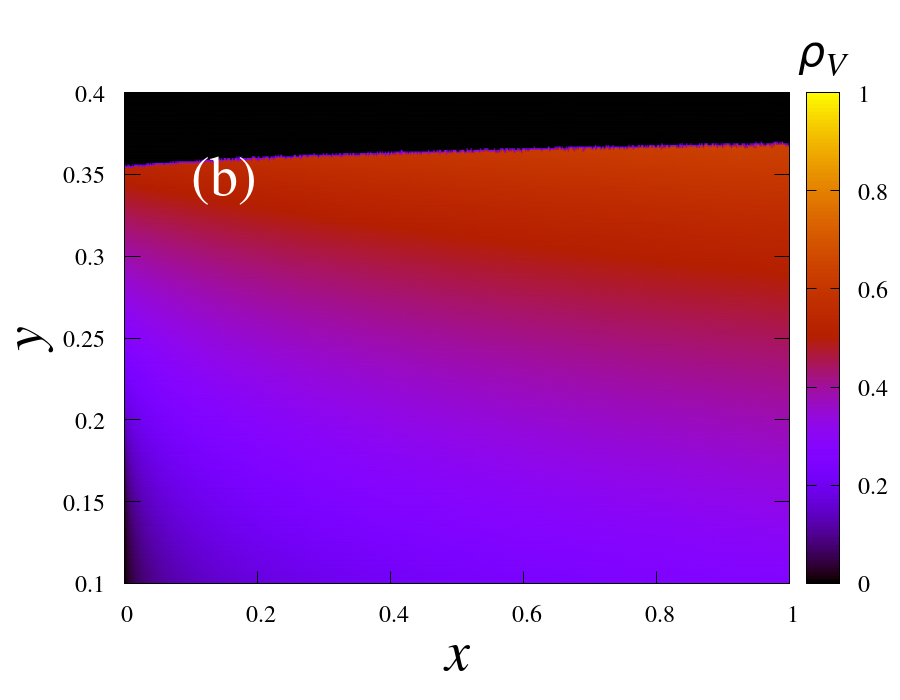}
\caption{Phase diagram of the YK model for (a) square and (b) hexagonal lattices for $r_{NO}=1.000$. Each figure shows $4\times 10^5$ points in $(x,\ y)$ space, each one representing the density of vacant sites, $\rho_V$. The black color stands for poisoned states, i.e., $\rho_V=0$.}
\label{fig:res_vaz_rno_1.000}
\end{center}
\end{figure}

Finally, we turn our attention to the phase diagram of the model for $r_{NO}=1.000$. Figure \ref{fig:res_vaz_rno_1.000} shows our results for both catalytic surfaces and, as can be seen, the continuous phase transition line exists only for very small values of $x$: for the square lattice (a), $x < 0.050$, and for the hexagonal one (b), $x < 0.030$.

\begin{figure} [!htbp] %h or !htbp
\begin{center}
\includegraphics[width = 3.1 in]{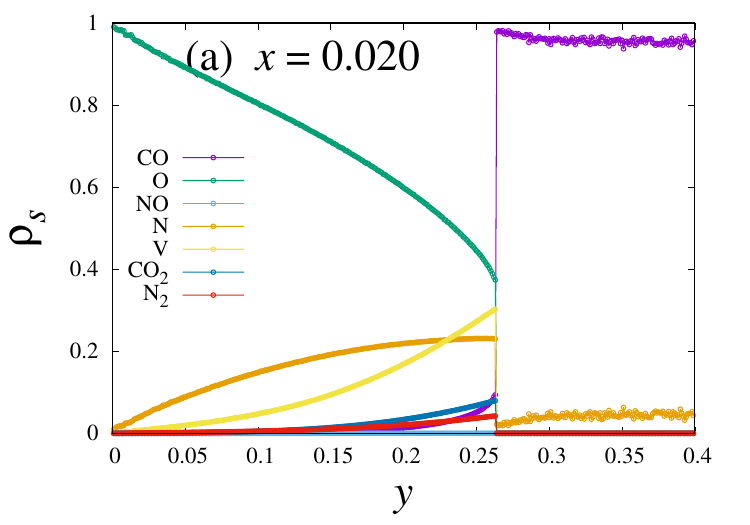}
\includegraphics[width = 3.1 in]{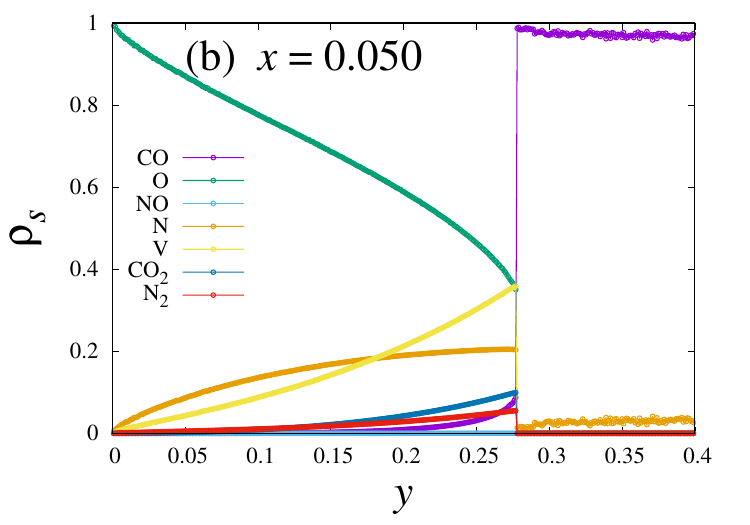}
\includegraphics[width = 3.1 in]{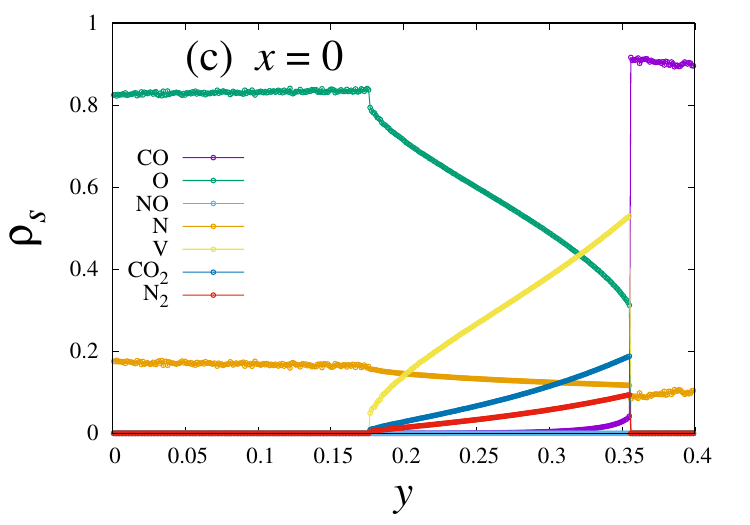}
\includegraphics[width = 3.1 in]{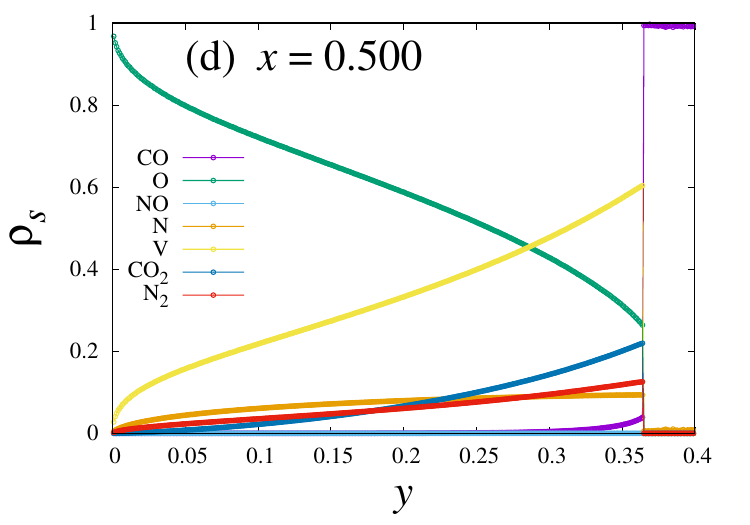}
\caption{Densities $\rho_s$ of the species presented in the YK model as function of the adsorption rate $y$ for square [(a) and (b)] and hexagonal [(c) and (d)] lattices for different values of $x$.}
\label{fig:snapshot}
\end{center}
\end{figure}

From these points on, the systems possess only one steady reactive state starting at $y=0.001$ and ending at the discontinuous phase transition line when it finally reaches the absorbing state, as can be seen in Fig. \ref{fig:snapshot} for both lattices.

Figure \ref{fig:snapshot}(a) and (b) presents some plots of the densities for $x=0.020$ and 0.050, respectively, for the square lattice, and Fig. \ref{fig:snapshot}(c) and (d) shows the curves for $x=0$ and 0.500, respectively, for the hexagonal lattice, showing that, different from the continuous phase transition, the discontinuous one is strong and is not destroyed by the inclusion of exchange diffusion to the model.

\section{Conclusions \label{conclusions}}

In this paper, we performed steady-state Monte Carlo simulations on the YK model in order to study the influence of the exchange diffusion on square and hexagonal lattices. In our simulations, CO molecules and N atoms are able to exchange places with each other or move to a neighboring vacant site at a given rate. We obtained a framework of the phase diagrams of the model for different values of the nitric oxide dissociation rate, $r_{NO}$, and showed the emergence of steady reactive state on square lattices and the appearance of the active phase for values smaller values of $r_{NO}$ on hexagonal ones. We also showed that the order of the transitions separating the absorbing states to the active phase remains unchanged when compared to the standard model, i.e., one continuous and other discontinuous, but $r_{NO}=1$ in which the continuous phase transition is destroyed.

Last, but not least, our results provide the first investigation of exchange diffusion of CO molecules and N atoms in the YK surface reaction model. Notably, even minimal diffusion induces the emergence of both continuous and discontinuous transitions on square lattices for some values of $r_{NO}$. For hexagonal lattices, where such transitions already occur in the absence of diffusion, the corresponding transition points shift as the diffusion rate increases, a behavior also observed for square lattices in this study.

\section*{Acknowledgments}

H.A. Fernandes, R. da Silva, and P.F. Gomes thank LaMCAD/UFG, the HPC resources of the IT Superintendency/USP, and the Lovelace cluster/UFRGS for computational support.

\section*{Funding Statement} 

This work was supported by the Brazilian agency CNPq (P. F. Gomes and H. A. Fernandes, grant no. 405508/2021-2; R. da Silva, grant no. 304575/2022-4) and by FAPEG (P. F. Gomes, grant no. 2019/10267000139).

%\section*{Data availability}

%All data generated in this research, along with the program codes and additional details, are available upon request from the authors: paulofreitasgomes@ufg.br, hafernandes@ufj.edu.br, and rdasilva@if.ufrgs.br.

\end{document}